\documentclass[twocolumn,showpacs,preprintnumbers,amsmath,amssymb]{revtex4}

\usepackage{graphicx}

\usepackage{graphicx}
\usepackage{dcolumn}
\usepackage{bm}
\usepackage{epsfig}
\usepackage{hyperref}

\begin{document}

\preprint{}

\title{Steady-state signatures of radiation trapping by cold multilevel atoms}

\author{Q. Baudouin} 
\author{N. Mercadier}
\author{R. Kaiser}
\affiliation{%
Universit\'e de Nice Sophia Antipolis, CNRS, Institut Non-Lin\'eaire de Nice, UMR 7335, F-06560 Valbonne, France\\
}%

\date{\today}

\begin{abstract}

In this paper, we use steady-state measurements to obtain evidence of radiation trapping in an optically thick a cloud of cold rubidium atoms. We investigate the fluorescence properties of our sample, pumped on opened transitions. This fluorescence exhibits a non trivial dependence on the optical thickness of the media. A simplified model, based on rate equations self-consistently coupled to a diffusive model of light transport, is used to explain the experimental observations in terms of incoherent radiation trapping on one spectral line. Measurements of the atomic populations and the fluorescence spectrum qualitatively agree with this interpretation. 

\end{abstract}


\maketitle

\section{Introduction}

Trapping of light due to multiple scattering by the atoms of a gas plays an important role in many transfer problems ranging from plasma physics to astrophysics \cite{Molisch1992}. Moreover, atomic vapors provide a well characterized set of identical and  very efficient resonant scatterers; from the beginning of the 20th century, they have been extensively used to experimentally investigate the phenomenon of radiation trapping \cite{Hayner1925, Zemansky1927, Kenty1932}. In hot atomic vapours, frequency redistribution due to atomic motion significantly affects
the radiation trapping process \cite{Holstein1947} leading to non diffusive multiple scattering processes \cite{Mercadier2009}.
From that prospect, samples of cold atoms, where frequency redistribution is strongly reduced, provide an interesting model experiment to study the role of multiple scattering with increasing complexity. Time resolved experiments on a closed atomic transition have demonstrated that cold atoms give raise to very efficient radiation trapping  \cite{FIORETTI, Labeyrie2003, Labeyrie2005}, consistent with steady state experiments on a closed transition \cite{Labeyrie2004}.
Even more subtle interference effects have been observed such as coherent backscattering \cite{Labeyrie1999} opening the way to study effects such localization of light in cold atomic vapours \cite{Akkermans2008}. 

Another additional interesting feature of such system is to provide light amplification when atomic vapours are optically pumped under certain conditions \cite{Mollow1972, Wu1977, Grison1991, Tabosa1991, Hilico1992a,Guerin2008}. For a given level of gain, when feedback due to multiple scattering is increased, a runaway regime can be reached where gain in the volume compensates for losses through the surface: then, the power emitted by the sample increases until saturation of gain is reached. This is the photonic bomb predicted by Letokhov\cite{Letokhov1968}. Above threshold, modal selection can occur, feedback due to multiple scattering favourising certain frequencies or directions of emission\cite{Tureci2008}. It has been predicted that gain and feedback due to multiple scattering can be combined to obtain random lasing action in a cloud of cold atoms\cite{Froufe-Perez2009}. 

In the context of quantum information, radiation trapping has been studied as a perturbative phenomena limiting the atomic coherence \cite{Stites2004}.
It has also been shown that radiation trapping inside the media can impact the photon statistics of light emitted outside the sample\cite{Beeler2003, Stites2004}. 
Finally, mechanical effects of trapped light also have been investigated and limits the density of atoms \cite{Sesko1991} or even induce mechanical instabilities \cite{Labeyrie2006, Mendonca2012}.

In this paper, we consider a configuration identified as a good candidate to obtain random lasing action, where multilevel atoms of rubidium are optically pumped on an open transition allowing for Raman gain to occur. We show that even in the absence of gain, light trapped inside the media can have a significant influence on the atomic populations, and therefore on the intensity and spectra of the emission of the sample. Understanding this regime is a required preliminary condition to later identify signatures of a random laser in this configuration. We stress that we exhibit here a model experiment allowing to investigate in laboratory environment features of radiation transfer in a multi-line system out of thermal equilibrium - well known in the domain of astrophysics \cite{Magnan1994,Lopez-Puertas2001,Faurobert2008}. 

The paper is organized as follows: section II describes the experimental setup, the atomic configuration under consideration and the main signatures of radiation trapping. In section III, we introduce a qualitative model (section III.A) based on a diffusion equation for light transport and rate equations for the atomic response to assess the influence of trapped light on the atomic emission. Detailed equations for the self-consistent solution of the coupled equations of the atomic response and a diffusion equation for light are presented in section III.B, together with a comparison to experimental results. Some approximations done in this simplified model are discussed in section III.C, before we conclude in section IV.

\section{Experiment}

\begin{figure}[t!]
\begin{center}
\includegraphics[width=0.45\textwidth]{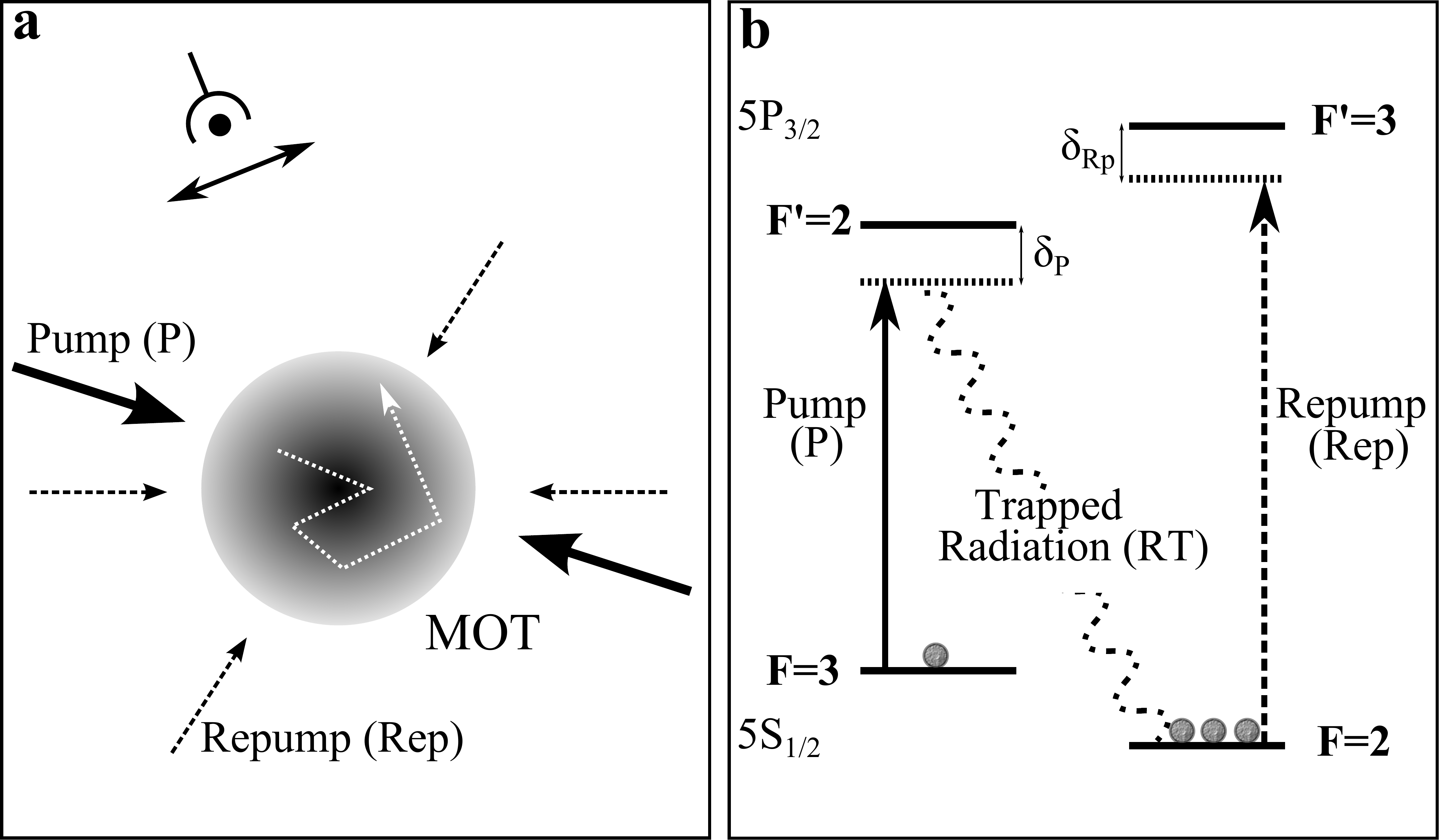}
\caption{Experimental setup. A cloud of $^{85}Rb$ is obtained in a magneto-optical trap, whose optical thickness can be adjusted by varying the dark MOT parameters, while keeping the number of atoms constant. A pump ($P$)  ($3\rightarrow  2'$ transition) and a repumper ($Rep$) ($2\rightarrow  3'$ transition) are then used to illuminate the atoms, while fluorescent emission from the sample is detected using a lens with a large solid angle and a high gain photodiode. }
\label{fig:setup}
\end{center}
\end{figure}

The main features of our magneto-optical trap have already been described in \cite{Guerin2010a}. We use 6 counter-propagating trapping beams with a waist of 3.4cm to load atoms of $^{85}$Rb from a vapor in a magneto-optical trap (see figure \ref{fig:setup}). Trapping beams are detuned by $-3 \Gamma$ from the $F=3 \rightarrow F'=4$ hyperfine transition of $^{85}$Rb, where $\Gamma$ is the width of the transition. The intensity of $\approx 3mW/cm^2$ per beam is slightly larger than the intensity saturation on the cycling transition ($I_{sat}=1.67 mW/cm^2$ \cite{Stecka}).  Six additional repumper beams tuned to the $F=2 \rightarrow F'=3$ transition maintain most of the atomic population in $F=3$.  We can load between $10^8$ and $10^{11}$ atoms by changing the background vapor pressure ($\sim 10^{-8}$ to a few $10^{-7}$ mbar) and the time used to load the trap ($\sim 10$ to $500$ ms). Once the atoms are loaded, we perform a temporal dark MOT by reducing the intensity of the repumper beams. This leads to a reduction of the size of the cloud and an increased spatial and optical density. By varying the duration of this compression process and with $10^9$ atoms initially loaded, it is possible to adjust the optical thickness $b_0$ of the sample ($b_0=-ln(T)$ where T is the coherent forward transmission measured on the $3\rightarrow 4'$ resonance) in a range of $20-75$. We stress that this protocol allows for changing the optical thickness while keeping the number of atoms quasi-constant ($\pm 10\%$). The RMS radius of the cloud are respectively 1.18mm and 0.61mm for $b_0=20$ and $b_0=75$. After this dark MOT, the trapping lasers and magnetic field gradients are switched off and we expose the sample to a pair of contrapropagating pump beams $P$ with a waist of $2.4$cm and a center intensity of $1.9$ mW/cm$^2$ per beam. The pump beam is obtained from a master laser which is then amplified by two stages of saturated slave lasers; hence, by tuning the frequency of the master laser using a double-pass acousto-optical modulator, we can scan the detuning $\delta_P$ of the pump with respect to the $F=3 \rightarrow F'=2$ transition by more than $16\Gamma$ without altering its intensity by more than $0.1\%$. Note that when the pump is detuned from the $F=3\rightarrow F'=2$ resonance, and if the $F=3$ state is more populated than $F=2$, Raman gain ($F=3 \rightarrow F'=2 \rightarrow F=2$) can be obtained in this system \cite{Kumar1985, Bowie2000,Grison1991, Tabosa1991, McKeever2003}. An additional  repumper on the $F=2 \rightarrow F'=3$ transition controls the population balance between the hyperfine ground states of $^{85}$Rb in the steady state regime. 
In the work described in this paper, we use 3 pairs of countra-propagating repumping beams with an intensity of $0.48mW/cm^2$ per beam, detuned by $-4\Gamma$ from the $F=2\rightarrow F'=3$ transition. Due to the low intensity of this repumper, most atoms are in the $F=2$ ground state when the pump is close to $F=3\rightarrow F'=2$ resonance. Note that the pump and repumper have respective waists of $2.4$ and $3$ $cm$, both much larger than the radius of our cloud ($\sim 1mm$). Hence, the mean intensities received by the atoms vary by less than $0.2\%$ in the whole range of compression ratios explored. 
Finally, fluorescent emission from our sample is detected in a solid angle of $0.01$str and measured using a high gain photodiode. Our measurements are made on a time scale of a few 100$\mu$s (typically 1 or 2 ms), while our system reaches its steady-state in less than a few $\mu$s even when radiation trapping occurs for the larger values of our optical thickness \cite{Labeyrie2005}. Hence, all observations reported here correspond to a steady-state regime. 

\begin{figure*}[htbp]
\begin{center}
\includegraphics[viewport = 0 0 400 350,width=0.7\textwidth]{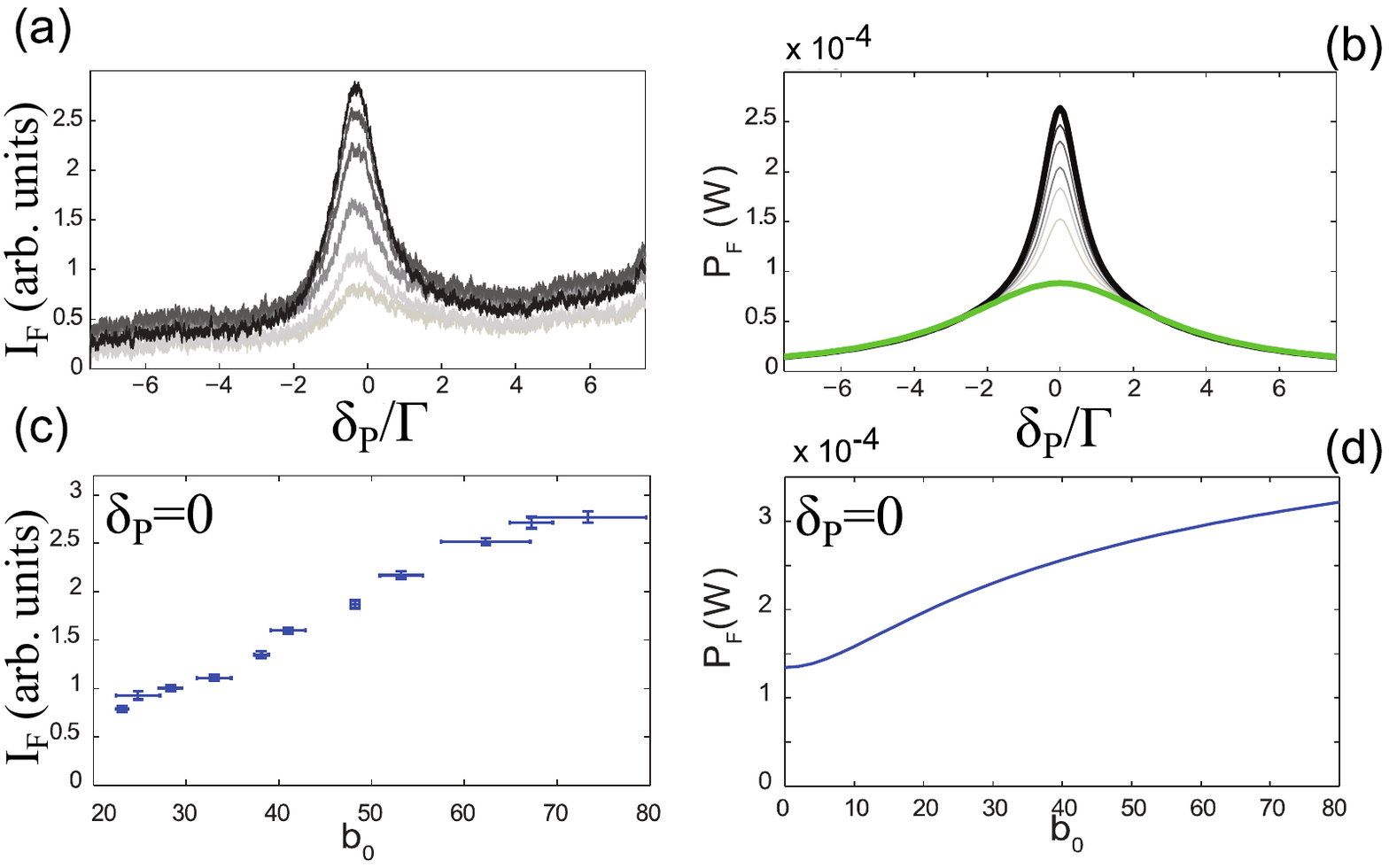}
\caption{(Color online) Evolution of the intensity of total fluorescence with the optical thickness $b_0$, for a sample of fixed number of atoms ($N=1.2\times 10^9$) . \textbf{a.} Experimental result of the total fluorescence, as a function of pump detuning, for various optical thicknesses $b_0$: from light to dark grey, 23, 33, 41, 53, 62 and 73. A clear increase of fluorescence for zero detuning of the pump is observed, attributed to radiation trapping effects. The intensity of the laser are : $1.88$mW/cm$^2$ for the pump laser and : $0.48$mW/cm$^2$ , for the repumper. The repumper is detuned by $-4 \Gamma$ from the $2-3'$ transition. \textbf{b.} Total power emitted by the sample in the same conditions, estimated from our model coupling diffusion and rate equations. \textbf{c.} Evolution with the optical thickness $b_0$ of the intensity of fluorescence measured experimentally for resonant pump ($\delta_P=0$). \textbf{d.} Corresponding total power emitted by the sample, estimated from our model, in good qualitative agreement with experimental observations. }
\label{fig:pic-central}
\end{center}
\end{figure*}

The most striking result of this experiment is shown on figure \ref{fig:pic-central}. We measure the fluorescence intensity of our sample submitted to a pump and to a weak and detuned repumper, as a function of the detuning of the pump. This detuning is scanned quickly, enough so that the cloud does not significantly expand or fall due to gravity, but slowly enough so that each point corresponds to a quasi steady-state regime. We have verified that dividing or multiplying by 2 this scanning rate does not affect our measurements. Furthermore, we have paid particular care to keep the number of atoms constant ($\pm 10\%$) when we change the optical thickness of the cloud. This protocol allows to easily distinguish  collective effects from a change of fluorescence due to an increased number of atoms.
If atoms would react individually to the exciting beams, the detected fluorescence would remain constant as we change the optical thickness $b_0$. As one can clearly see in figure \ref{fig:pic-central}, we observe a strong increase in this intensity with $b_0$ when the pump is tuned to the $F=3\rightarrow F'=2$ resonance. This is a signature of a strong collective emission effect that occurs in our sample.

\begin{figure}[t]
\begin{center}
\includegraphics[viewport = 0 0 350 200,width=0.7\textwidth]{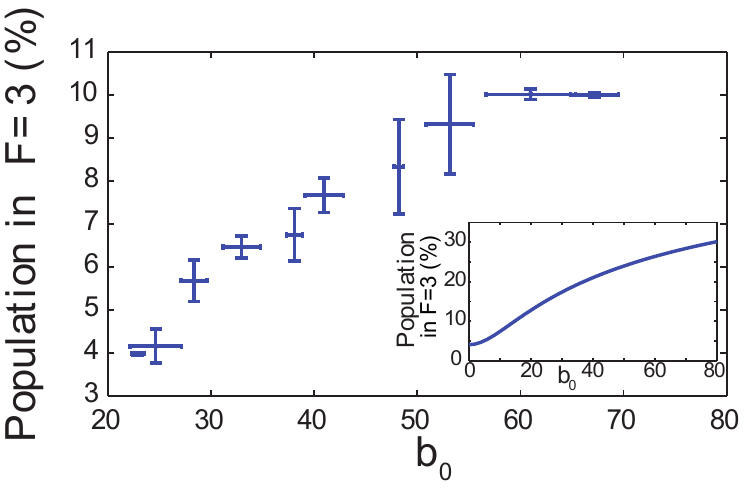}
\caption{(Color online) Experimental measurements of the atomic population in the F=3 ground state, as a function of the optical thickness $b_0$, obtained from absorption imaging. Pumping conditions are the same as for figure \ref{fig:pic-central}, and the pump is on resonance. A clear increase of the population of $F=3$ with the optical thickness is observed. \textbf{Inset} : Evolution with $b_0$ of the population of the $F=3$ state predicted by our model. We note the overestimation of this population for large optical thickness, indicating limitations of our model.}
\label{fig:populations}
\end{center}
\end{figure}

\begin{figure}[t]
\begin{center}
\includegraphics[viewport = 0 0 350 200,width=0.7\textwidth]{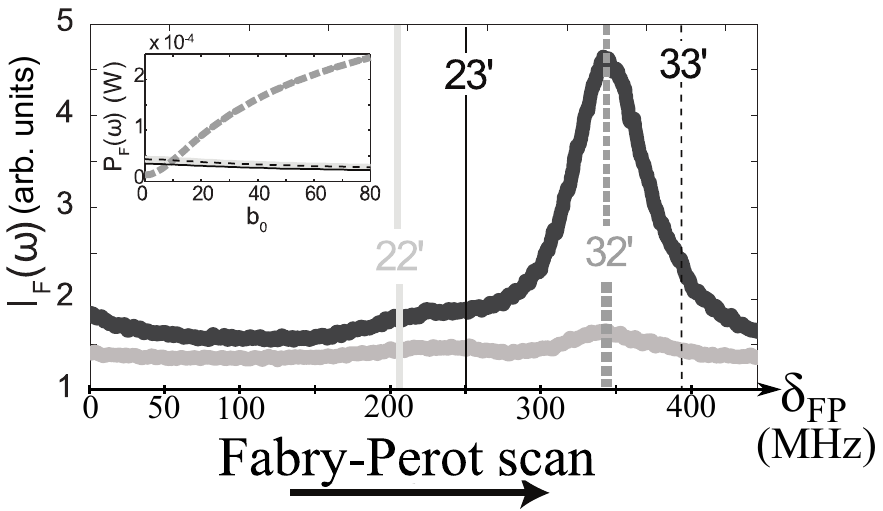}
\caption{Experimental fluorescence spectrum of the emitted line, obtained by scanning the length of a Fabry-Perot cavity, for an on resonance optical thickness of 50 (light grey) and 145 (dark grey).  Pumping conditions are identical to those of figure \ref{fig:pic-central}, but the number of atoms has been increased to $1.4~~10^{10}$ to maximize the signal at the output of the cavity. A clear increase with the optical thickness of the intensity emitted on the $F=3 \rightarrow F'=2$ line is observed. Note that $F=2 \rightarrow F'=2$ and $F=2 \rightarrow F'=3$ lines are separated from by the $F=3 \rightarrow F'=2$ and $F=3\rightarrow F'=3$ lines by $\approx 3 $GHz, and thus correspond to different longitudinal modes of our Fabry-Perot interferometer, even if they appear almost superimposed in this scan. \textbf{Inset:} Evolution of the intensity emitted on each hyperfine line of $^{85}Rb$ D2 line as a function of the optical thickness, estimated with our model. A slight decrease of all lines is predicted, excpet the line corresponding to the $F=3 \rightarrow F'=2$ transition (thick dotted grey line), compatible with experimental observations.}
\label{fig:spectres}
\end{center}
\end{figure}

\section{Self-consistent Model of atomic response coupled to Radiation transfer of light}
\subsection{Qualitative explanation}

In what follows, we will show how these observations can be understood by taking into account the strong impact of diffuse light on the atomic populations. A qualitative description allows to understand the dominant phenomenon of this situation and involves a description of the atomic response and a diffusion equation for radiation trapping of the light on one particular atomic emission line. 

The atomic response can be obtained using optical Bloch equations, describing atomic populations and (optical and Zeeman) coherences. For a two level transition excited with a single optical frequency, steady state results for populations and emission intensities can be equivalently obtained from simpler rate equations. Note that a similar simplification can be used for time dependant quantities if the decay rate of atomic coherences is much larger than those of populations, a situation which can e.g. occur in presence of collisions.
The steady state solution of the rate equations is given by $\rho_{ee}=\frac{B(\delta) I}{B(\delta) I +\Gamma}\rho_{gg}$, where $\rho_{ee}$, $\rho_{gg}$ are respectively excited and ground population , with $B$ the Einstein coefficient  (see details below), $I$ the incident laser intensity and $\delta$ the detuning between the laser frequency and the atomic transition frequency.

When radiation trapping effects can be neglected, our atomic four level scheme (figure \ref{fig:setup}) reduces to two 2-level atoms excited each by a laser, allowing to use rate equations for the steady state populations. These two 2-level systems are coupled by incoherent radiative decay (from $F'=2 \rightarrow F=2$ and $F'=3 \rightarrow F=3$).

In the experiments described in this paper we focus on a situation with a strong pump laser tuned close to the atomic line $F=3 \rightarrow F'=2$ and a weaker repumper laser detuned from the atomic line $F=2 \rightarrow F'=3$. In this case most atoms are in the ground state $F=2$.
We measure the total fluorescence $P_{F}$ of the atomic cloud, which is proportional to the total population $\rho_{2'2'}+\rho_{3'3'}$  of the excited state of states $F'=2$ and $F'=3$ :

\begin{align}
\label{eq:fluo}
P_{F} &\propto \rho_{2'2'}+\rho_{3'3'} 
  \\    &\propto \rho_{33}\frac{B_{32'}(\delta_P) I_P}{\Gamma+B_{32'}(\delta_P) I_P}+\rho_{22} \frac{B_{23'}(\delta_{Rep}) I_{Rep}}{\Gamma+B_{23'}(\delta_{Rep}) I_{Rep}},
\end{align}

with $\delta_P$ (resp. $\delta_{rep}$) the detuning of pump (resp. repumper) laser, and $I_P$ (resp. $I_{rep}$) the corresponding intensities. The populations of the ground states $F=2$ ($F=3$) are denoted $\rho_{22}$ ($\rho_{33}$). For the parameters of our experiment described in this paper and with the Einstein coefficients $B_{ij}$ defined in eq (\ref{bij}) below, we have $B_{32'}(\delta_P=0) I_P=0.061/\Gamma$ and $B_{23'}(\delta_{Rep}=-4\Gamma) I_{Rep}=0.0027/\Gamma$, so that we can approximate the above expression by: 

\begin{equation}
\label{eq:fluo2}
P_{F} \propto  \rho_{33}B_{32'}(\delta_P) I_P+\rho_{22} B_{23'}(\delta_{Rep}) I_{Rep}
\end{equation}

In the weak driving limit where ($B_{32'}(\delta_P) I_P \ll \Gamma, B_{23'}(\delta_{Rep}) I_{Rep} \ll \Gamma$), most of the population is in the ground state $\rho_{33}+\rho_{22} \approx 1$. One can now understand, that in the regime where $B_{32'}(\delta_P) I_P > B_{23'}(\delta_{Rep})I_{Rep} $, a transfer of populations from the $F=2$ state to the $F=3$ state will increase the total emission. The specific feature of the work presented in this paper is the role of the multiple scattered light along the $F=2\rightarrow F'=2$ transition, which acts as en effectif additional repumper laser, transfering more atoms from the $F=2$ into the $F=3$ hyperfine level. 
For low optical thickness, when atoms from the $F=3$ level are transfered into the $F=2$ level via a spontaneous Raman scattering process, the scattered photon can easily escape the cloud. However for larger optical thickness, this spontaneous Raman photon can be reabsorbed by one of the many atoms in the $F=2$ level which is thus pumped into the $F=3$ level. The depumping of the strong pump beam is thus reduced and more atoms will end up in the $F=3$ level, leading to an increased total fluorescence of the cloud.

Additional measurements confirm this scenario. Using standard absorption imaging, we can measure the fraction of atoms in the $F=3$ state after quickly switching off the pump and repump lasers. As shown on figure \ref{fig:populations}, a clear increase of the population of the $F=3$ state with the optical thickness is observed. Another signature is obtained by measuring the optical spectrum of the emitted light. Using a Fabry-Perot cavity with a free spectral range of $400$MHz and a very low finesse of 14, we have measured the spectrum of the emitted light by coupling part of the quasi-isotropic emission from the sample to the Fabry-Perot cavity. With a resolution of $\approx 30$MHz the different hyperfines lines of $^{85}Rb$ can be resolved. As one can see in figure \ref{fig:spectres}, we observe a strong increase of the emission on the $F=3 \rightarrow F'=2$ state  for larger optical thickness of the sample. Both these results confirm the qualitative explanation given above: the increase of fluorescence intensity with $b_0$ is explained by the enhanced repumping of atoms into the $F=3$ ground state.

\subsection{Detailed description of the model}

To go beyond this qualitative description and to assess to what extend the population redistribution due to the diffuse light in the sample 
can explain the observed effects, we will now turn to a more quantitative description based on an ab initio model of radiation trapping in the steady state regime. 

In what follows, we assume that only light with a frequency close to the $F=2 \rightarrow F'=2$ transition can be trapped in the system and influence the atomic population balance. This assumption is reasonable at least for intermediate optical thickness, since light scattering induced by the repumper, both on the $F=2 \rightarrow F'=3$ and the $F=3 \rightarrow F'=3$ line, is detuned by several line widths $\Gamma$ from any resonance ($\delta_{Rep}=-4\Gamma$). Furthermore pump induced light scattering on the $F=3 \rightarrow F'=2$ transition is subject to low optical thickness, as the $F=3$ state is weakly populated (at least before population redistribution occurs due to radiation trapping). To first order, radiation trapping will thus occur for light scattered by the atoms at a frequency close to the $F=2 \rightarrow F'=2$ resonance. Its effects on atoms can be modeled by an incoherent pumping rate. Finally, we only consider 4 effective nondegenerate hyperfine levels and do not take into account the complex Zeeman substructure of  the different hyperfine levels. We thus neglect effects induced by the polarisation of the scattered light \cite{Faurobert2008} or the possible existence of dark states due to effects of coherence between different Zeeman sublevels. 

Since only two independant and non degenerate transitions are coherently excited (the effect on the atoms of the trapped light being described by an incoherent pump rate), the stationary optical Bloch equations reduce to stationary rate equations:
\begin{equation}
\label{eq:rate}
\begin{split}
\dot{\rho}_{22}=0   = &- B_{23'}(\delta_{Rep}) I_{Rep} (\rho_{22}-\rho_{3'3'})\\ &- B_{22'}(\delta_{RT}) I_{RT} (\rho_{22}-\rho_{2'2'})\\ &+ \Gamma (T_{3'2} \rho_{3'3'} + T_{2'2} \rho_{2'2'})\\
\dot{\rho}_{33}=0   = &- B_{32'}(\delta_{P}) I_{P}(\rho_{33}-\rho_{2'2'}) \\ &+\Gamma (T_{3'3} \rho_{3'3'} + T_{2'3} \rho_{2'2'})\\
\dot{\rho}_{2'2'}=0 = &B_{32'}(\delta_{P}) I_{P} (\rho_{33}-\rho_{2'2'})\\ &+B_{22'}(\delta_{RT}) I_{RT} (\rho_{22}-\rho_{2'2'})- \Gamma \rho_{2'2'} \\
\dot{\rho}_{3'3'}=0 = &B_{23'}(\delta_{Rep}) I_{Rep}(\rho_{22}-\rho_{3'3'}) - \Gamma \rho_{3'3'} \\
1=&\rho_{2'2'}+\rho_{3'3'}+\rho_{22}+\rho_{33}
\end{split}
\end{equation}

where $\rho_{ii}$ are the populations of state i and $T_{i'j}$ the branching ratios for desexcitation from state $i'$ to state $j$ (accounting for the degeneracy of states $i'$ and $j$ assuming a statistical population in all Zeeman sublevels). $I_{P}$, $I_{Rep}$, $I_{RT}$ are respectively the intensities of the pump, repumper, and diffuse light, and $\delta_P$, $\delta_{Rep}$, $\delta_{RT}$ their respective detuning to the transitions they excite. Finally, the detuning dependant Einstein coefficients $B_{ij}$ can be written as :
\begin{equation}
B_{ij}(\delta) = \frac{1}{3}\frac{2F'+1}{2F+1}T_{ij}\frac{ \sigma_0}{ \hbar \omega_0} \frac{1}{1 + 4\frac{\delta^2}{\Gamma^2}}
\label{bij}
\end{equation}
where $\sigma_0=3\lambda_0^2/2\pi$ is the on resonance cross section of a two level atom with a transition frequency $\omega_0=2\pi c/\lambda_0$. For our level scheme, the values of $T_{ij}=\frac{2J'+1}{2J+1}\frac{2F+1}{2F'+1}S_{F'F}$  are : $T_{2'2}=1-T_{2'3}=14/18$ and $T_{3'3}=1-T_{3'2}=10/18$, where $S_{F'F}$ are the strength factors of the transitions\cite{Stecka}.

As the optical thickness at frequencies of the external pump and repumper lasers is small, the external laser beams are weakly attenuated and the corresponding intensities $I_P$ and $I_{Rep}$ can be considered homogeneous. On the other hand, the intensity $I_{RT}$ of diffuse light at $\omega\approx\omega_{22'}$ is not known a priori and can strongly vary in space. In the stationary regime, we compute its profile by solving a diffusion equation:

\begin{equation}
\label{eq:diffusion}
  \frac{1}{3 n \sigma_{ext}(\mathbf{r})} \Delta I_{RT} (\mathbf{r}) = - n W(\mathbf{r}) 
\end{equation}

where $\mathbf{r}$ is the position in the cloud. $I_{RT} (\mathbf{r})$ and $\sigma_{ext}(\mathbf{r})$ are respectively the spatially dependant intensity of the diffuse light and 
the extinction cross section. The source term $W(\mathbf{r})$ on the right hand side of this diffusion equation describes the emission of photons from the excited state $F'=2$ and takes into account the reabsorption of light at the frequency $\omega_{22'}$. We note that using the prefactor $1/(3 n \sigma_{ext}(\mathbf{r}))$ in equation \ref{eq:diffusion} implies that the effect of absorption on the diffusion process has been neglected \cite{Elaloufi2003, Pierrat2006a}.
Indeed, in contrast to the situation where the emitted light can be understood as the scattering of an external laser, we do not have an incident laser on the $F=2'\rightarrow F=2$ line. The energy of the diffuse light is partly taken from the light scattered by other atoms, requiring to count this as a loss in the energy balance term $W(\mathbf{r})$.
In the case of a two level atom without pump, emitted power exactly balances the extinction and a diffusion equation without source term is the adequate description.

We stress that the extinction cross section and the source term  depend on the atomic populations : 
\begin{equation}
\label{eq:crosssection}
\begin{split}
& \sigma_{ext}(\mathbf{r}) = \frac{2F'+1}{2F+1}\frac{T_{2'2} \sigma_0}{1 + 4 (\delta_P/\Gamma)^2} (\rho_{2'2'} (\mathbf{r}) - \rho_{22} (\mathbf{r})) \\ 
& W(\mathbf{r}) = \hbar \omega_0 \Gamma T_{22'} \rho_{22'}(\mathbf{r}) - \sigma_{ext}(\mathbf{r}) I_{RT}(\mathbf{r})
\end{split}
\end{equation} 

The coupling between the evolution of the atomic populations (eqs. \ref{eq:rate}) and the trapped light (\ref{eq:diffusion}) is now explicit and we will use a self-consistent solution of these equations. This coupling implies that $I_{RT}$, $W$ and $\sigma_{ext}$ are all connected and also depend of $b_0$.
We also assume the spatial density of atoms  $n$ to have a homogeneous spherical symmetry. The density of atoms $n$ is thus linked to the optical thickness of the cloud by:
\begin{equation}
n=\frac{b_0^{3/2} \sqrt{\pi}}{\sigma_0^{3/2}\sqrt{6N}}
\end{equation}
where $N$ is the total number of atoms, which is kept constant. Equation \ref{eq:diffusion} thus reduces to :
\begin{equation}
\label{eq:diffusion2}
{  \frac{2 N \sigma_0^3}{\pi \sigma_{ext}(\mathbf{r})} \Delta I_{RT} (\mathbf{r}) = - b_0^3 W(\mathbf{r})}
\end{equation}

To simplify the algebra, we consider a spherical cloud, so that the problem becomes rotation invariant, and the diffusion equation reduces to a one dimensional equation in spherical coordinates. The coupled eqs. \ref{eq:rate} and \ref{eq:diffusion} are solved in a self-consistent way, using an iterative process and a classical Runge-Kutta method for Eq. \ref{eq:diffusion}. 

We now turn to the evaluation of experimentally observable quantities. The most convenient signal to be measured in our experiment is the total power emitted by the cloud. Energy conservation implies that, in the steady state regime, the total scattered power has to be taken from the incident lasers beams, ie pump and repumper. The total power of fluorescence can thus be written as :
\begin{equation}
P_F =  \int n(r,b_0) 4 \pi r^2 dr [ \sigma_{ext}^P(r,b_0) I_{P}(r)+\sigma_{ext}^{Rep}(r,b_0) I_{Rep}(r)]
\end{equation}
where $\sigma_{ext}^P(r)$ and $\sigma_{ext}^{Rep}(r)$ are the extinction cross sections of the pump and repumper beams respectively. In our case, the pump and repumper are weakly attenuated, and the spatial dependance of their intensity can be neglected. In the following we will omit indicating the dependence on $b_0$. These cross sections can be obtained using the results of the optical Bloch equations, as done for the extinction cross section of the light around the $F=2 \rightarrow F'=2$ line (see eq. \ref{eq:crosssection}). It is however possible to show that this emitted power $P_F$ can also be written as :

\begin{equation}
P_F =  \int  n(r) 4 \pi r^2 dr [ \hbar \omega_0\Gamma (\rho_{2'2'}(r)   + \rho_{3'3'}(r))- \sigma_{ext}(r) I_{RT}(r)] \nonumber
\label{eq:fluoRT}
\end{equation}

This expression highlights that the total power emitted is the sum of the powers emitted by each single atom on each transition, balanced by the power re-absorbed by the cloud. We can also compute the power emitted on each transition using a similar approach. In figure \ref{fig:pic-central} we plot the value of the total emitted power evaluated for a cloud of constant density. 
Considering the number of approximations, the model yields a satisfactory agreement with our experimental observations.

Using these self-consistent solutions, we can also derive other quantities which can be measured in the experiment, as for instance the population of the hyperfine ground states. The qualitative explanation described in section (III.A) of the impact of radiation trapping on the emission of our cloud has been based on the increase of the population in $F=3$ for increasing optical thickness. 
Cold atom experiments provide the possibility to measure the populations of the hyperfine ground states after swichting off all incident laser beams, the small excited state population quickly relaxing to the ground states. The experimental result and the theoretical prediction are shown in figure \ref{fig:populations}, illustrating the very satisfactory agreement between the predictions of our model and the experimental results. 

A further quantity which can be computed and confronted to an experimental verification is the power emitted along each of the hyperfine spectral lines.
From the populations of the excited states and the various branching ratios for the emission, we can derive the emission on all relevant lines. We note that for the line where radiation trapping is present ($F=2\rightarrow F'=2$) we take into account reabsorption via the $- \sigma_{ext}(r) I_{RT}(r)$ term (see (\ref{eq:fluoRT}), whereas such reabsorption has been neglected for the other spectral lines.
The experimental observation of the spectrum of the emitted light in cold atom experiments is technically more challenging, but with a moderate resolution we have been able to resolve the different lines of the relevant transitions. Figure \ref{fig:spectres} shows the experimental and the numerical results, in agreement with our qualitative description, since an increased fluorescence along the $F=3 \rightarrow F'=2$ line is observed for the larger value of $b_0$.

\subsection{Beyond Rate Equations for the atomic response}

Our simple model, using stationary rate equations coupled with one diffusion equation, agrees qualitatively with experimental observations. Despite this satisfactory result, several limitations of our models might account for quantitative differences observed between numerical and experimental results.
We have for instance neglected the Zeeman substructure of all hyperfine levels, polarization effects and Raman scattering among different Zeeman sublevels are not taken into account. As the incident lasers are polarized, we expect some degree of polarization to remain, at least for moderate radiation trapping. Thus, a more refined model similar to those used in astrophsyics \cite{Faurobert2008} would probably allow to improve the description of the light emitted by the cloud of cold atoms and measurements of the polarization along the various emission lines will allow to test the regime of validity of such more evolved codes for radiation trapping.

Another assumption made in the present work has been to consider radiation trapping along one single optical line. 
This allows to use one diffusion equation self-consistently coupled to a model of the atomic response. Extending this model to take into account radiation trapping along other spectral lines is in principle possible. 
In the regime of parameters for which we have performed the experiments however, the use of a single diffusion equation is justified by the important differences in the optical thicknesses associated to the atomic transitions. Neglecting saturation effects, we can e.g. estimate the optical thickness along each line by 
\begin{equation}
\begin{split}
& b^{32'} \approx \frac{b_0 \rho_{33}}{1+4 \frac{\delta_P^2}{\Gamma^2}}=b_0 \rho_{33} \\ 
& b^{22'} \approx \frac{b_0 \rho_{22}}{1+4 \frac{\delta_{P}^2}{\Gamma^2}}=b_0 \rho_{22} \\
& b^{23'} \approx \frac{b_0 \rho_{22}}{1+4 \frac{\delta_{Rep}^2}{\Gamma^2}}=\frac{b_0 }{65} \rho_{22}\\ 
& b^{33'} \approx \frac{b_0 \rho_{33}}{1+4 \frac{\delta_{Rep}^2}{\Gamma^2}}=\frac{b_0 }{65} \rho_{33}, \\
\end{split}
\end{equation} 
where $b^{ij'}$ is the optical thickness of the transition between $F=i$ and $F'=j$.
When $b^{ij'} \gg 1$ ($L \gg l_{sc}$) radiation trapping on the corresponding line has to be taken into account in our model.
According to experimental results shown on figure \ref{fig:populations} and for the lower values of $b_0$ ($b_0=23$), $\rho_{33}\approx 4\%$ so that the respective line optical thicknesses are: $b^{32'} \approx 1$, $b^{22'} \approx 23$, $b^{23'} \approx 0.35$ and  $b^{33'} \approx 0.015$. In this regime, the dominant radiation trapping occurs on the 
$F=2 \rightarrow F'=2$ transition for which trapping has been taken into account.
For larger values of $b_0$ however ($b_0=68$), $\rho_{33}\approx 10\%$ and $b^{32'} \approx 7$, $b^{22'} \approx 61$, $b^{23'} \approx 1$ and  $b^{33'} \approx 0.1$. Given the optical thickness on the pump line ($F=3 \rightarrow F'=2$), multiple scattering on this line would in principle have to be taken into account, even if saturation effects on this transition are expected to reduce the impact of such radiation trapping. Note that saturation of the atomic transition is properly described in our self-consistent model, as long as rate equations are valid.

\begin{figure}[t]

\begin{center}
\includegraphics[viewport = 0 0 400 150,width=0.7\textwidth]{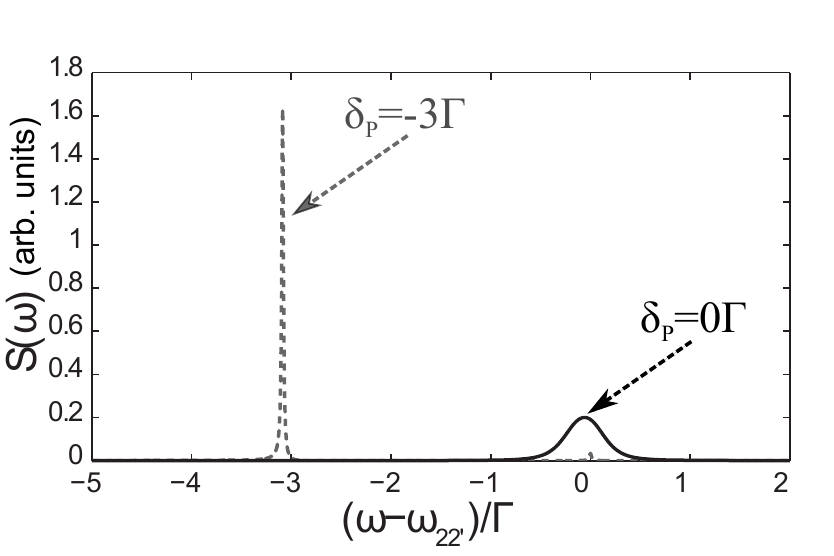}
\caption{Computed emission spectrum close to the $2\rightarrow2'$ transition for an isolated atom, using optical Bloch equations and the quantum regression theorem. The Rabi frequency of 
repumper is $\Omega_{Rep}=2 \Gamma$ and its detuning  is $\delta_{Rep} = -5\Gamma$. The Rabi frequency of the pump laser is  $\Omega_{P}=2 \Gamma$  and its detuning $\delta_P=0$ (continuous line) and $\delta_P=-3\Gamma$ (dotted line). While the atomic emission spectrum has a width comparable to the transition width when the pump is resonant, it becomes significantly narrower when it is detuned by a few $\Gamma$. }
\label{fig:spectra}
\end{center}
\end{figure}

Another assumption made in the approach we have used in this work concerns the coherence properties of diffuse light. 
We have used rate equations to describe the light-atom coupling. This approach neglects in particular any atomic coherence between different (hyperfine or Zeeman) ground states. In order to evaluate the possibility of such ground state coherences, we have computed the optical spectrum of the light emitted along the $F=2 \rightarrow F'=2$ transition, applying the quantum regression theorem~\cite{Lax1968, MercadierPhD} to atoms with several hyperfine levels (neglecting the Zeeman structure).
As shown in figure \ref{fig:spectra} the spectrum of light emitted by an isolated atom driven by a pump close to resonance has a width of the order of the natural linewidth $\Gamma$. The increase of the total fluorescence as a function of the optical thickness thus occurs in a regime where the linewidth of the scattered light is of the order of the width of the excited state, making the use of an incoherent radiation trapping model a reasonable approach. However when the pump laser is detuned further away from resonance, the emission line can become significantly narrower that the natural width of the transition. We therefore expect Raman coherences between the different hyperfine ground states to play a more prominent role when the pump laser will be detuned from the $F=3 \rightarrow F'=2$ transition, allowing even hyperfine Raman gain to appear. This situation is explored in our work on random lasing with cold atoms where gain and scattering need to be combined.

\section{Conclusion}

In this paper, we have demonstrated that radiation trapping in a cloud of cold atoms can significantly alter the emission properties of atomic clouds. Simple and strong evidence of radiation trapping can be thus obtained in a steady-state regime, in contrast to studies exploiting trapping times \cite{Holstein1947,FIORETTI, Labeyrie2003, Labeyrie2005}. Cold atom clouds thus appear as excellent candidates to investigate in a laboratory different regimes of radiation trapping out of thermal equilibrium, characterized by a strong coupling between radiation and atomic populations. In the case considered in this work, a simple model coupling in a self-consistent way a diffusion equation (describing light transport) and rate equations (describing  the atomic behavior) has allowed to explain qualitatively all our observations.

\begin{acknowledgements}
We acknowledge financial support from the program ANR-06-BLAN-0096 and funding for N.M. and Q.B. by DGA. We thank V. Guarrera for her contribution to the experimental setup and W. Guerin for careful reading of this paper.
\end{acknowledgements}


\begin{thebibliography}{10}



\bibitem{Molisch1992}
A. Molisch, B. Oehry, and G. Magerl, Journal of Quantitative Spectroscopy and
  Radiative Transfer {\bf 48},  377  (1992).

\bibitem{Hayner1925}
L. Hayner, Physical Review {\bf 26},  364  (1925).

\bibitem{Zemansky1927}
M.~W. Zemansky, Physical Review {\bf 29},  513  (1927).

\bibitem{Kenty1932}
C. Kenty, Physical Review {\bf 42},  823  (1932).

\bibitem{Holstein1947}
T. Holstein, Physical Review {\bf 72},  1212  (1947).

\bibitem{Mercadier2009}
N. Mercadier, W. Guerin, M. Chevrollier, and R. Kaiser, Nature Physics {\bf 5},
   602  (2009).

\bibitem{FIORETTI}
A. Fioretti {\it et~al.}, Optics communications {\bf 149},  415  (1998).

\bibitem{Labeyrie2003}
G. Labeyrie {\it et~al.}, Physical Review Letters {\bf 91},  223904  (2003).

\bibitem{Labeyrie2005}
G. Labeyrie, R. Kaiser, and D. Delande, Applied Physics B {\bf 81},  1001
  (2005).

\bibitem{Labeyrie2004}
G. Labeyrie {\it et~al.}, Optics Communications {\bf 243},  157  (2004).

\bibitem{Labeyrie1999}
G. Labeyrie {\it et~al.}, Physical Review Letters {\bf 83},  5266  (1999).

\bibitem{Akkermans2008}
E. Akkermans, A. Gero, and R. Kaiser, Physical Review Letters {\bf 101},  103602
  (2008).

\bibitem{Mollow1972}
B.~R. Mollow, Physical Review A {\bf 5},  2217  (1972).

\bibitem{Wu1977}
F. Y. Wu, S. Ezekiel, M. Ducloy, and B.R. Mollow, Physical Review Letters {\bf 38},
  1077  (1977).

\bibitem{Grison1991}
D. Grison {\it et~al.}, Europhysics Letters (EPL) {\bf 15},  149  (1991).

\bibitem{Tabosa1991}
J.W.R. Tabosa, G. Chen, Z. Hu, R.B. Lee, H.J. Kimble, Physical Review Letters {\bf 66},  3245  (1991).

\bibitem{Hilico1992a}
L. Hilico, C. Fabre, and E. Giacobino, Europhysics Letters (EPL) {\bf 18},  685
   (1992).

\bibitem{Guerin2008}
W. Guerin, F. Michaud, and R. Kaiser, Physical Review Letters {\bf 101},  093002
  (2008).

\bibitem{Letokhov1968}
V.~S. Letokhov, Soviet Journal of Experimental and Theoretical Physics {\bf
  26},  835  (1968).

\bibitem{Tureci2008}
H.~E. T\"{u}reci, L. Ge, S. Rotter, and a.~D. Stone, Science (New York, N.Y.)
  {\bf 320},  643  (2008).

\bibitem{Froufe-Perez2009}
L.S. Froufe-P\'{e}rez, W. Guerin, R. Carminati, and R. Kaiser, Physical Review
  Letters {\bf 102},  173903  (2009).

\bibitem{Stites2004}
R. Stites {\it et~al.}, Optics letters {\bf 29},  2713  (2004).

\bibitem{Beeler2003}
M. Beeler, R. Stites, S. Kim, L. Feeney, S. Bali, Physical Review A {\bf 68},  013411  (2003).

\bibitem{Sesko1991}
D.~W. Sesko, T.~G. Walker, and C.~E. Wieman, Journal of the Optical Society of
  America B {\bf 8},  946  (1991).

\bibitem{Labeyrie2006}
G. Labeyrie, F. Michaud, and R. Kaiser, Physical Review Letters {\bf 96},
  023003  (2006).

\bibitem{Mendonca2012} T. Mendonca and R. Kaiser, Physical Review Letters {\bf 108} ,033001 (2012), 


\bibitem{Magnan1994}
C. Magnan and P. {De Laverny}, Astrophysics {\bf 37},  167  (1994).

\bibitem{Lopez-Puertas2001}
M. Lopez-Puertas and F.~W. Taylor, {\em {Non LTE radiative transfer in the
  atmosphere}} (World Scientific, ADDRESS, 2001), p.\ 487.

\bibitem{Faurobert2008}
M. Faurobert, M. Derouich, V. Bommier, and J. Arnaud, Astronomy {\bf 206},  201
   (2008).

\bibitem{Guerin2010a}
W. Guerin {\it et~al.}, Journal of Optics {\bf 12},  24002  (2010).

\bibitem{Kumar1985}
P. Kumar and J.~H. Shapiro, Optics Letters {\bf 10},  226  (1985).

\bibitem{Bowie2000}
J.L. Bowie, J.C. Garrison, and R.Y. Chiao, Physical Review A {\bf 61},  053811
  (2000).

\bibitem{McKeever2003}
J. McKeever {\it et~al.}, Nature {\bf 425},  268  (2003).

\bibitem{Stecka}
D.~A. Steck, {Rubidium 85 D Line Data, http://steck.us/alkalidata}, 2008.

\bibitem{Elaloufi2003}
R. Elaloufi, R. Carminati, and J.~J. Greffet, JOSA A {\bf 20},  678  (2003).

\bibitem{Pierrat2006a}
R. Pierrat, J.~J. Greffet, and R. Carminati, JOSA A {\bf 23},  1106  (2006).

\bibitem{Lax1968}
M. Lax, Physical Review {\bf 172},  350  (1968).

\bibitem{MercadierPhD}
N. Mercadier, Ph.D. thesis, Universit\'{e} de Nice Sophia-Antipolis, 2011.

\end{thebibliography}
\end{document}